\documentclass[aps,prl,twocolumn,floats]{revtex4}

\usepackage{amssymb}
\usepackage{amsmath}
\usepackage{graphicx}
\usepackage{bm}
\usepackage{mdframed}
\usepackage{color}
\usepackage{gensymb}
\usepackage{comment}

\begin{document}

\bibliographystyle{prsty}

\title{Dynamically stabilized spin superfluidity in frustrated magnets}

\author{Ricardo Zarzuela,$^{1}$ Daniel Hill,$^{2}$ Jairo Sinova$^{1,3}$ and Yaroslav Tserkovnyak$^{2}$}

\affiliation{$^{1}$ Institut f\"{u}r Physik, Johannes Gutenberg Universit\"{a}t Mainz, D-55099 Mainz, Germany \\ 
$^{2}$ Department of Physics and Astronomy, University of California, Los Angeles, California 90095, USA \\
$^{3}$ Institute of Physics Academy of Sciences of the Czech Republic, Cukrovarnick\'{a} 10, 162 00 Praha 6, Czech Republic
}

\begin{abstract}
We study the onset of spin superfluidity, namely coherent spin transport mediated by a topological spin texture, in frustrated exchange-dominated magnetic systems, engendered by an external magnetic field. We show that for typical device geometries used in nonlocal magnetotransport experiments, the magnetic field stabilizes a spin superflow against fluctuations, up to a critical current. For a given current, the critical field depends on the precessional frequency of the texture, which can be separately controlled. We contrast such dynamic stabilization of a spin superfluid to the conventional approaches based on topological stabilization. 
\end{abstract}
\maketitle

\textit{Introduction.}| Low-dissipation spin-dependent transport offers pro\- mising perspectives for the design of high-speed communication architectures in the next generation of spintronic devices \cite{Upadhyaya-PRL2017,Tserkovnyak-PRL2017}. A prominent example lies in the concept of spin superfluidity \cite{Sonin-SSC1978,Sonin-JETP1978,Sonin-JETP1979,Konig-PRL2001,Sonin-AiP2010}, introduced in the late 70s by analogy with the phenomena of (charge) superconductivity and mass superfluidity in $^{4}$He, albeit not exempt from dissipation in the form of spin relaxation/dephasing. One of the hallmarks of spin superfluidity is the algebraic decay of spin signals over long distances, in contrast to the exponential suppression observed for (incoherent) magnon-mediated transport \cite{Takei-PRL2014,Takei-PRB2014}. Experimental signatures of superfluid spin transport have been recently reported in (high-quality) antiferromagnetic platforms for device geometries usually used in nonlocal magnetotransport measurements \cite{Yuan-SciAdv2018,Stepanov-NatPhys2018}. However, parasitic anisotropies arising naturally in the fabrication process of spintronic (collinear) magnetic systems have a detrimental effect on the spin superfluid state, since these break the underlying spin O(2) symmetry of the (anti)ferromagnetic host and, therefore, open a gap in the excitation spectrum that lifts the Goldstone mode sustaining the spin superflow. 

Magnetic systems with frustrated interactions dominated by isotropic exchange offer an alternative route to overcome these key challenges, by averaging the parasitic anisotropies out at the macroscopic level and restoring an effective SO(3) symmetry in the spin space. Furthermore, the minimal description of these noncollinear magnetic platforms is provided by the O(4)-nonlinear sigma model \cite{Dombre-PRB1989,Azaria-PRL1992,Chubukov-PRL1994}, whose excitation spectrum for bulk systems consists of three Goldstone branches. An external magnetic field lifts two of these soft (angular) modes \cite{Tserkovnyak-PRB2017}. The remaining linearly dispersing mode (corresponding to rotations along the direction of the magnetic field), in turn, is able to sustain a phase-coherent precessional state, which can be triggered by (interfacial) spin-orbit torques \cite{Ochoa-PRB2018}. In particular, for the nonlocal device geometries usually considered (i.e., thin films with an out-of-plane orientation of the spin accumulations and the external magnetic field), the collective flapping out of the basal plane initiates the unwinding (phase slips) of the order parameter. Therefore, gapping these angular modes by an external magnetic field should impede the (topological) relaxation channel for the spin superflow, enhancing its stability.

It is, therefore, vital to understand to what extent the fluctuations of the SO(3)-valued order parameter are detrimental to a collective spin flow in frustrated magnets, in the presence of the stabilizing effect of the magnetic field. In this Letter, we show that the in-plane (rotational) Goldstone mode can sustain spin supercurrents and study the corresponding steady-state solution. We examine the robustness of the underlying phase-coherent spin configuration against collective fluctuations and determine its stability threshold as a function of the magnetic field and the precessional frequency. In this regard, we show that even though, under purely topological considerations, there is a low-energy route for the collective relaxation of the spin-superfluid state, the analysis of its dynamics dictates a finite range of stable solutions. It is, therefore, not topology, which is dictated by the free energy (as, e.g., in the case of easy-plane anisotropies  \cite{Ochoa-PRB2018}), that is key here but a field-induced dynamic stabilization of the basal winding textures.

\textit{Equations of motion.}| At the macroscopic level, frustrated magnets are described by an SO(3) order parameter, $\hat{R}(\vec{r},t)$, which represents smooth and slowly varying rotations of the initial spin configuration \cite{Dombre-PRB1989,Halperin-PRB1977}. Dynamics of $\hat{R}$ and the nonequilibrium spin density $\bm{m}(\vec{r},t)$ of the system are governed by the equations \cite{Ochoa-PRB2018,Tserkovnyak-PRB2017}:
\begin{eqnarray}
\label{EoM1}
\bm{m}&=&\chi(\bm{\omega}+\gamma\bm{B}),\\
\label{EoM2}
\partial_{t}\bm{m}+\alpha s\bm{\omega}&=&\mathcal{A}\nabla\cdot\vec{\bm{\Omega}}+\gamma\,\bm{m}\bm{\times}\bm{B},
\end{eqnarray}
where $\chi$, $\gamma$ and $\mathcal{A}$ denote the spin susceptibility, the gyromagnetic ratio and the order-parameter stiffness, respectively. Furthermore, $\alpha$ parametrizes losses due to dissipative processes (Gilbert damping) in the bulk and $s\simeq\hbar S/a^{3}$, with $S$ and $a$ being the length of the microscopic spin operators and the lattice spacing, respectively. $\bm{B}$ denotes the external magnetic field, $\bm{\omega}\equiv i\textrm{Tr}\left[\hat{R}^{\top}\hat{\bm{L}}\partial_{t}\hat{R}\right]/2$ is the local precessional frequency and the spin fields $\bm{\Omega}_{k}\equiv i\textrm{Tr}\left[\hat{R}^{\top}\hat{\bm{L}}\partial_{k}\hat{R}\right]/2$, $k=x,y,z$, describe the spatial variations of the instantaneous state of the spin texture, with $[\hat{L}_{\alpha}]_{\beta\gamma}\equiv-i\epsilon_{\alpha\beta\gamma}$ being the generators of the SO(3) group. Note that the dissipative term $\alpha s\bm{\omega}$ in Eq.~\eqref{EoM2} is the one responsible for the algebraic decay of the spin signal when balanced with the appropriate boundary conditions for the spin supercurrent, as we will show in the next section.

A convenient representation of the order parameter is given in terms of unit-norm quaternions, $\mathbf{q}=(w,\bm{v})$ \cite{Ochoa-PRB2018}. The parametrization $w=\cos(\phi/2)$ and $\bm{v}=\sin(\phi/2)\,\bm{n}$, where $\bm{n}$ and $\phi(\vec{r},t)$ represent the rotation axis and the local rotation angle for spins, respectively, yields the well-known Rodrigues' rotation formula: $\hat{R}_{\alpha\beta}=\cos\phi\,\delta_{\alpha\beta}+(1-\cos\phi)n_{\alpha}n_{\beta}+\sin\phi\,\epsilon_{\alpha\gamma\beta} n_{\gamma}$. By applying this representation to Eqs.~\eqref{EoM1} and \eqref{EoM2}, we obtain the following equation of motion for the quaternion \cite{SM}:
\begin{align}
\label{EoM_q}
&\partial_{t}^{2}\mathbf{q}+\frac{1}{T}\partial_{t}\mathbf{q}-v_{m}^{2}\nabla^{2}\mathbf{q}+\lambda(\mathbf{q})\mathbf{q}\\
&\hspace{1.9cm}-\gamma\left(\mathbf{B}^{\star}\hspace{-0.1cm}\wedge\partial_{t}\mathbf{q}+\tfrac{1}{2}\partial_{t}\mathbf{B}^{\star}\hspace{-0.1cm}\wedge\mathbf{q}\right)=0,\nonumber
\end{align}
where $T\equiv \chi/\alpha s$ defines a characteristic relaxation time, $v_{m}\equiv\sqrt{\mathcal{A}/\chi}$ denotes the speed of spin waves in the magnetic medium and $\lambda(\mathbf{q})\equiv\partial_{t}\mathbf{q}\bm{\odot}\partial_{t}\mathbf{q}^{\star}-v_{m}^{2}\partial_{k}\mathbf{q}\bm{\odot}\partial_{k}\mathbf{q}^{\star}+\tfrac{\gamma}{2}\bm{\omega}\bm{\cdot}\bm{B}$ is a quadratic prefactor. Here, $\mathbf{q}^{\star}\equiv(w,-\bm{v})$ denotes the adjoint quaternion, $\mathbf{B}\equiv(0,\bm{B})$, the Hamilton product is defined by $\mathbf{q}_{1}\wedge\mathbf{q}_{2}\equiv(w_{1}w_{2}-\bm{v}_{1}\bm{\cdot}\bm{v}_{2},w_{1}\bm{v}_{2}+w_{2}\bm{v}_{1}+\bm{v}_{1}\bm{\times}\bm{v}_{2})$, and the scalar product reads $\mathbf{q}_{1}\bm{\odot}\mathbf{q}_{2}\equiv\tfrac{1}{2}\left(\mathbf{q}_{1}\wedge\mathbf{q}_{2}^{\star}+\mathbf{q}_{2}\wedge\mathbf{q}_{1}^{\star}\right)=w_{1}w_{2}+\bm{v}_{1}\bm{\cdot}\bm{v}_{2}$.

\begin{figure}[h!]
\begin{center}
\includegraphics[width=1.0\columnwidth]{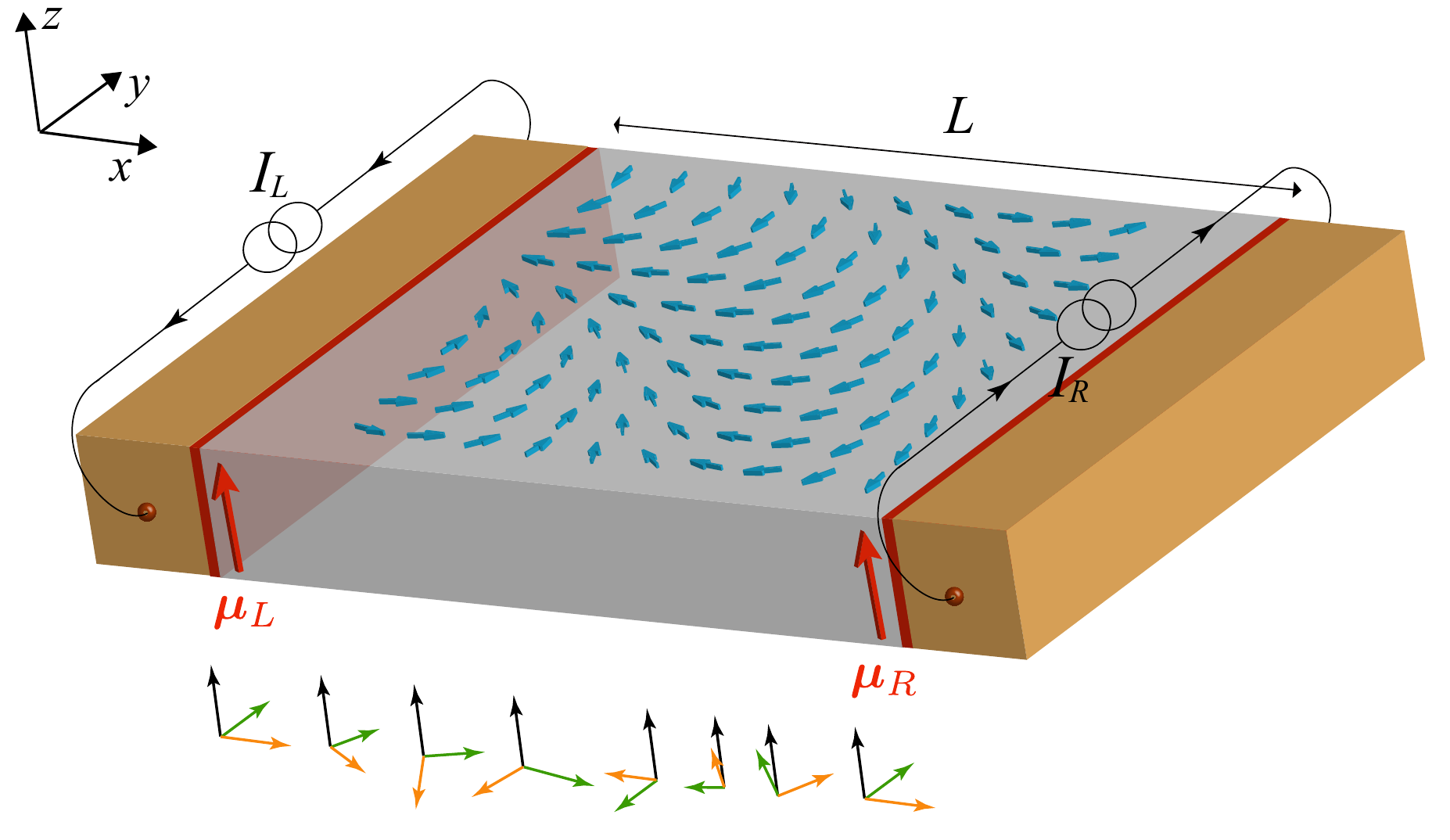}
\caption{Two-terminal geometry for the generation and detection of spin superfluidity in frustrated magnetic platforms. The spin precession (blue arrows) along the spin accumulations $\bm{\mu}_{L,R}$ (red arrows) is depicted as a rotating triad of vectors, which represents the internal spin frame of the texture. The black arrow represents the rotation axis $\bm{n}$ and the local rotation angle $\phi$ is illustrated by the rotation of the green and orange arrows of each triad within the plane perpendicular to $\bm{n}$ (which is oriented along the $z$ axis here).}
\vspace{-0.5cm} 
\label{Fig1}
\end{center}
\end{figure}

\textit{Spin superfluid state.}| Hereafter, we restrict ourselves to a quasi-one-dimensional geometry, by assuming translational symmetry along the $y$ and $z$ spatial directions, and a finite length $L$ along the $x$ direction, see Fig.~\ref{Fig1}. We also consider lateral contacts extending along the $yz$ plane for spin injection via spin Hall physics \cite{Sinova-RMP2015} and assume that the spin accumulations $\bm{\mu}_{L}$ (left interface) and $\bm{\mu}_{R}$ (right interface) are parallel to the $z$ axis, which set the (uniform) direction of $\bm{n}=\hat{\bm{e}}_{z}$ across the sample. Furthermore, we apply the magnetic field collinear as well, $\bm{B}=B\bm{n}$, to stabilize the superflow, and take $\phi(\vec{r},t)$ to be spatially smooth and slowly varying. As a result, Eq.~\eqref{EoM_q} turns into the following dynamical equation for the rotation angle \cite{SM}:
\begin{equation}
\label{EoM_phi}
\partial_{t}^{2}\phi+\frac{1}{T}\partial_{t}\phi-v_{m}^{2}\partial_{x}^{2}\phi+\gamma\partial_{t}B=0.
\end{equation}
The precessional steady solution to the above equation is obtained by considering the ansatz $\phi(x,t)=X(x)+\tau(t)$. The boundary conditions are
\begin{eqnarray}
\label{BC1}
-2\mathcal{A}\,\partial_{x}\mathbf{q}\wedge\mathbf{q}^{\star}|_{L}&=&\frac{g_{L}}{4\pi}\big[\bm{\mu}_{L}-2\hbar\partial_{t}\mathbf{q}\wedge\mathbf{q}^{\star}|_{L}\big],\\
\label{BC2}
2\mathcal{A}\,\partial_{x}\mathbf{q}\wedge\mathbf{q}^{\star}|_{R}&=&\frac{g_{R}}{4\pi}\big[\bm{\mu}_{R}-2\hbar\partial_{t}\mathbf{q}\wedge\mathbf{q}^{\star}|_{R}\big],
\end{eqnarray}
which arise from the exchange of angular momentum between the magnet and adjacent (heavy-)metal contacts in the form of ordinary exchange torques and enhanced Gilbert damping \cite{Ochoa-PRB2018,Tserkovnyak-PRB2017}. Here, $g_{L}$ and $g_{R}$ denote the spin mixing conductance at the left ($x=-\tfrac{L}{2}$) and right ($x=\tfrac{L}{2}$) terminals, respectively \cite{FN2}. In what follows, we assume the same spin mixing conductance at both interfaces, $g\equiv g_{L}=g_{R}$. The spin superfluid state is therefore described, under an external magnetic field, by \cite{SM}
\begin{equation}
\label{SS_sol}
\phi_{s}(x,t)=\phi_{0}+kx+\frac{1}{2}\left(\frac{\alpha s}{\mathcal{A}}\right)\omega x^{2}+\omega t,
\end{equation}
with
\begin{eqnarray}
\label{k}
k&=&\frac{g}{8\pi\mathcal{A}}\left(\mu_{R}-\mu_{L}\right),\\
\label{omega}
\omega&=&\frac{g}{2}\frac{\mu_{L}+\mu_{R}}{\hbar g+2\pi\alpha s L}.
\end{eqnarray}
We therefore conclude that the external magnetic field has no effect on the spin texture sustaining the spin superflow \cite{Takei-PRL2014,Takei-PRB2014}. 

\textit{Fluctuations and stability.}| We proceed next to study the robustness of the spin superfluid state, Eq.~\eqref{SS_sol}, against fluctuations of the order parameter. First, we introduce the following orthonormal set $\{\mathbf{q}_{s},\bm{\xi}_{1},\bm{\xi}_{2},\bm{\xi}_{3}\}$ of quaternions, where $\mathbf{q}_{s}$ corresponds to the superfluid solution given by Eq.~\eqref{SS_sol}, $\bm{\xi}_{1}=\left(0,\hat{\bm{e}}_{y}\right)$, $\bm{\xi}_{2}=\left(0,\hat{\bm{e}}_{x}\right)$, and $\bm{\xi}_{3}=2\partial_{\phi_{s}}\mathbf{q}_{s}=\big(\hspace{-0.1cm}-\sin(\phi_{s}/2),\cos(\phi_{s}/2)\hat{\bm{e}}_{z}\big)$. Note that $\bm{\xi}_{1,2}$ represent $\pi$ rotations around the $y$ and $x$ axes, respectively. We incorporate fluctuations into the order parameter via the following parametrization:
\begin{equation}
\label{op_fluct}
\mathbf{q}=\mathbf{q}_{s}\sqrt{1-2|\Psi_{t}|^{2}-\Psi_{l}^{2}}+\hat{\bm{e}}_{+}\Psi_{t}+\hat{\bm{e}}_{-}\Psi_{t}^{*}+\bm{\xi}_{3}\Psi_{l},
\end{equation}
where $\Psi_{t}$ (complex valued) and $\Psi_{l}$ (real valued) represent the transverse and longitudinal (with respect to the rotation axis) fluctuation modes of the (unit-norm) quaternion order parameter. Here, $\hat{\bm{e}}_{\pm}\equiv\tfrac{1}{\sqrt{2}}(\bm{\xi}_{1}\pm i\bm{\xi}_{2})$ are chiral quaternions with the properties $\hat{\bm{e}}_{\pm}^{2}=0$ and $\hat{\bm{e}}_{+}\bm{\odot}\hat{\bm{e}}_{-}=\hat{\bm{e}}_{-}\bm{\odot}\hat{\bm{e}}_{+}=1$. We note that the transverse fluctuation modes $\Psi_{t}$ correspond to the out-of-plane rotations along the $x$- and $y$-axis, which are gapped out by the magnetic field. By inserting this parametrization into Eq.~\eqref{EoM_q} and keeping terms up to first order in $\Psi_{t}$ and $\Psi_{l}$, we obtain the following dynamical equations for the fluctuation fields \cite{SM}:
\begin{align}
\label{EoM_l}
&\partial_{t}^{2}\Psi_{l}+\frac{1}{T}\partial_{t}\Psi_{l}-v_{m}^{2}\partial_{x}^{2}\Psi_{l}=0,\\
\label{EoM_t}
&\partial_{t}^{2}\Psi_{t}+\left(\frac{1}{T}-i\gamma B\right)\partial_{t}\Psi_{t}-v_{m}^{2}\partial_{x}^{2}\Psi_{t}\\
&\hspace{0.8cm}+\frac{1}{4}\Big[(\partial_{t}\phi_{s})^{2}-v_{m}^{2}(\partial_{x}\phi_{s})^{2}+2\gamma B\partial_{t}\phi_{s}\Big]\Psi_{t}=0.\nonumber
\end{align}
Stability analysis of Eq.~\eqref{EoM_l} in Fourier space (with respect to the $x$ coordinate) yields the eigenvalues
\begin{equation}
\label{eigen_l}
\lambda_{l}^{\pm}(q)=\frac{1}{2T}\left[\pm\sqrt{1-\frac{q^{2}}{q_{c}^{2}}}-1\right],
\end{equation}
where $q$ and $q_{c}\equiv1/2v_{m}T$ denote the Fourier wavevector and its critical value, respectively. For $|q|\leq q_{c}$, the eigenvalues of the dynamical system are real valued and negative, $\lambda_{l}^{\pm}\leq0$. For $|q|>q_{c}$ the eigenvalues are complex valued, $\lambda_{l}^{\pm}(q)=-\tfrac{1}{2T}\pm i\Omega(q)$. Consequently, $\bar{\Psi}_{l}\propto e^{\lambda_{l}^{\pm}t}\sim e^{-t/2T}$ decays exponentially with time, so that the spin superfluid solution \eqref{SS_sol} is robust against longitudinal fluctuations. This is in agreement with the fact that rotations within the basal ($xy$) plane cannot unwind the order parameter.

\begin{figure}[t!]
\begin{center}
\includegraphics[width=1.0\columnwidth]{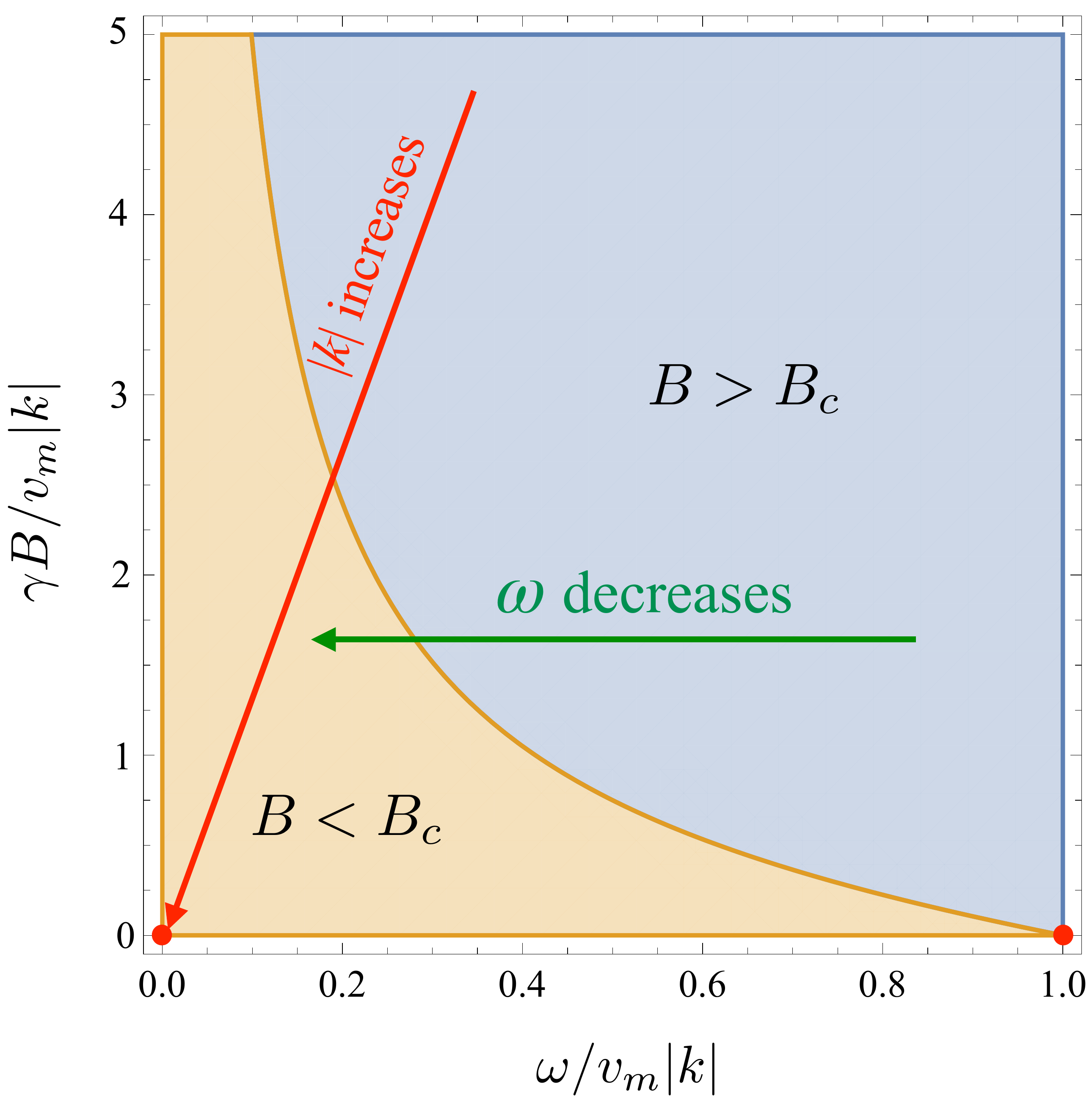}
\caption{Stability diagram and critical threshold for the spin superfluid state \eqref{SS_sol} parametrized by the normalized Larmor (magnetic field) and precessional ($\omega$) frequencies. The phase-coherent precessional state is stable for magnetic fields above the critical one (blue region). Red dots indicate the two special points of the diagram, namely $(0,0)$ (unstable) and $(0,1)$ (stable). Red and green lines display the paths within the diagram parametrized by the increase of the wavevector $|k|$ (for constant $B$ and $\omega$) and the decrease of $\omega$ (for fixed $B$ and $|k|$), respectively.}
\vspace{-0.5cm} 
\label{Fig2}
\end{center}
\end{figure}

In the case of transverse fluctuations, we suppose that the winding of the superfluid phase, described by $\partial_{x}\phi_{s}$, is changing smoothly across the magnet. As a result, we perform a local stability analysis of the $\Psi_{t}$ modes supposing $\partial_{x}\phi_{s}\simeq k$ is approximately constant. Again, by Fourier transforming Eq.~\eqref{EoM_t} we obtain the dynamical system
\begin{align}
\label{FEoM_t}
&\partial_{t}^{2}\bar{\Psi}_{t}+\left(\frac{1}{T}-i\gamma B\right)\partial_{t}\bar{\Psi}_{t}\\
&\hspace{1.5cm}+\left[\tfrac{\omega^{2}}{4}+v_{m}^{2}\big(q^{2}-\tfrac{k^{2}}{4}\big)+\tfrac{\gamma}{2}\omega B\right]\bar{\Psi}_{t}=0,\nonumber
\end{align}
where $\bar{\psi}(q,t)\equiv\mathcal{F}[\psi(x,t)]$ denotes the Fourier transform of the fluctuation field. Stability analysis of the above dynamical system yields the eigenvalues
\begin{equation}
\label{eigen_t}
\lambda_{t}^{\pm}=\frac{1}{2}\left\{-\frac{1}{T}\pm\textrm{Re}(Z)+i\left[\gamma B\pm\textrm{Im}(Z)\right]\right\},
\end{equation}
with $Z\equiv\sqrt{(i\gamma B-\tfrac{1}{T})^{2}-\left(\omega^{2}+v_{m}^{2}(4q^{2}-k^{2})+2\gamma\omega B\right)}$. The instability threshold arises when $-\tfrac{1}{T}\pm\textrm{Re}(Z)>0$, since the transverse fluctuation will therefore blow up as time increases.  Working out this condition leads to the following critical value for the magnetic field, above which the phase-coherent precessional state \eqref{SS_sol} is robust against fluctuations of the order parameter:
\begin{equation}
\label{critical_B}
2\gamma B_{c}=\frac{v_{m}^{2}k^{2}}{\omega}-\omega,
\end{equation}
which holds up to the frequency $\omega_{c}=v_{m}|k|$. Figure~\ref{Fig2} depicts the stability diagram of the spin superfluid state \eqref{SS_sol} in terms of the precessional frequency and the external magnetic field.

\textit{Discussion.}| From Eq.~\eqref{SS_sol} we can clearly extract two length scales for the spin-superfluid phase, namely $\ell_{\textrm{p}}=2\pi/|k|$ and $\ell_{\textrm{nl}}=\sqrt{\mathcal{A}/\alpha s\omega}$. The former determines the pitch of the magnetic spiral sustaining the spin-carrying state, whereas the latter determines the spatial rate of change of the associated wave number (due to damping). Our constant-pitch treatment for the instability, based on the local wave number $k$, should work when $k$ does not vary much on the lengthscale set by fluctuation wavelengths $q^{-1}$. This translates into the condition $\ell^{-1}_{\textrm{nl}}\ll\sqrt{kq}$. It thus needs to be verified, for internal consistency, that the corresponding lower bound for $q$ (which becomes progressively smaller as $\alpha\to0$) does not significantly affect the critical magnetic field \cite{SM}. Let us now discuss in some detail the physics encapsulated in the corresponding local-stability diagram, Fig.~\ref{Fig2}.

First, there are two special points in Fig.~\ref{Fig2}, namely $\mathbf{p}_{u}=(0,0)$ and $\mathbf{p}_{s}=(0,1)$. The former is always unstable, where phase slips are triggered by flapping out of the basal plane. On the other hand, $\mathbf{p}_{s}$ is stable, since here the phase velocity associated with the spin superflow becomes faster than the spin waves and the superfluid cannot relax towards the uniform magnetic configuration. Second, keeping $\omega$ and $B$ fixed, by increasing the wave vector $|k|$ we move along the straight path (red line) that converges asymptotically towards $\mathbf{p}_{u}$, leading to a stability-instability transition. This means, in particular, that when we have a single spin-current injector (e.g., $\bm{\mu}_{L}\neq\bm{0}$ and $\bm{\mu}_{R}=\bm{0}$ in the Fig.~\ref{Fig1} setup), the local wave vector $k(x)$ increases towards the injector, where the instability thus sets in first. Third, by fixing $B$ and $|k|$ and decreasing the frequency ($\omega\rightarrow0$) we move along the horizontal green line towards the unstable regime. This case is relevant, for example, when we increases the channel length $L$ and/or Gilbert damping $\alpha$, which would reduce the frequency, according to Eq.~\eqref{omega}.

A question that naturally arises is what happens beyond the stability regime. Since the onset of instability occurs close to the injector, we speculate that special boundary solutions (such as the contact-soliton ones found in conventional ferromagnetic platforms \cite{Iacocca-PRB2019,Schneider-2018}) or chaotic dynamics may emerge near the injector, which would suppress the overall spin-current injection. A lower but finite spin current may then still propagate along the channel, once the stable regime is reached at a low enough wave vector. We leave it as an open question to elucidate how the unstable boundary dynamics settle into a stable steady flow in the bulk. Finally, as discussed in Ref.~\onlinecite{Ochoa-PRB2018}, the degradation of the spin superflow in the absence of magnetic field and low frequencies occurs via phase slippage mediated by (Anderson-Toulouse) $4\pi$ vortices \cite{Anderson-PRL1976}. As a result, a mesoscopic residual spin current may prevail, $\propto|\nabla\phi_{s}|\simeq 4\pi/L$, which becomes negligible in macroscopic samples. A large enough magnetic field precludes the nucleation of these topological textures by gapping out the out-of-plane rotations.

\acknowledgements

This research was supported in part by the National Science Foundation under Grant No. NSF PHY-1748958. RZ and JS acknowledge support by the Transregional Collaborative Research Center (SFB/TRR) 173 SPIN+X, the Dynamics and Topology Centre funded by the State of Rhineland Palatinate and the Alexander von Humboldt Foundation. YT is supported by the NSF under Grant No. DMR-1742928 and the Alexander von Humboldt Foundation and is grateful for the hospitality of the University of Mainz, where most of this work has been carried out.

\newpage

\onecolumngrid
 
\section*{\Large Supplemental Material}

\section{Derivation of equation (3)}

To begin with, we incorporate Eq.~\eqref{EoM1} into Eq.~\eqref{EoM2} and thus obtain the equation
\begin{equation}
\label{eq1}
\partial_{t}\bm{\omega}=v_{m}^{2}\nabla\cdot\vec{\bm{\Omega}}+\gamma(\bm{\omega}\bm{\times}\bm{B}-\partial_{t}\bm{B})-\frac{1}{T}\bm{\omega}.
\end{equation}
Second, in the quaternion representation the identities $\bm{\omega}=2\partial_{t}\mathbf{q}\wedge\mathbf{q}^{\star}$ and $\bm{\Omega}_{k}=2\partial_{k}\mathbf{q}\wedge\mathbf{q}^{\star}$, $k=x,y,z$ hold. Third, we apply the Hamilton product $\mathbf{q}^{\star}\wedge\cdot$ to Eq.~\eqref{eq1}, so that we obtain
\begin{equation}
\label{eq2}
\partial_{t}^{2}\mathbf{q}^{\star}+\frac{1}{T}\partial_{t}\mathbf{q}^{\star}-v_{m}^{2}\nabla^{2}\mathbf{q}^{\star}+\lambda(\mathbf{q})\mathbf{q}^{\star}-\gamma\left(\partial_{t}\mathbf{q}^{\star}\hspace{-0.1cm}\wedge\mathbf{B}+\tfrac{1}{2}\bm{q}^{\star}\hspace{-0.1cm}\wedge\partial_{t}\mathbf{B}\right)=0,\nonumber
\end{equation}
taking into account the expressions
\begin{align}
\mathbf{q}^{\star}\wedge\partial_{\mu}(\partial_{\mu}\mathbf{q}\wedge\mathbf{q}^{\star})&=-\partial_{\mu}^{2}\mathbf{q}^{\star}-(\partial_{\mu}\mathbf{q}\wedge\partial_{\mu}\mathbf{q}^{\star})\mathbf{q}^{\star},\\
\mathbf{q}^{\star}\wedge(\partial_{\mu}\mathbf{q}\wedge\mathbf{q}^{\star})&=-\partial_{\mu}\mathbf{q}^{\star},\\
\mathbf{q}^{\star}\wedge[(\partial_{\mu}\mathbf{q}\wedge\mathbf{q}^{\star})\bm{\times}\bm{B}]&=-\partial_{\mu}\mathbf{q}^{\star}\wedge\mathbf{B}+\frac{1}{2}(\bm{\omega}\bm{\cdot}\bm{B})\mathbf{q}^{\star}.
\end{align}
Finally, by applying the adjoint operator to the above equation and using the identity $(\mathbf{q}_{1}\wedge\mathbf{q}_{2})^{\star}=\mathbf{q}_{2}^{\star}\wedge\mathbf{q}_{1}^{\star}$, we derive Eq.~\eqref{EoM_q}.

\section{Phase-coherent precessional solution to equation (7)}

We start by noting that, if $\phi_{s}(x,t)$ denotes the precessional angle of the spin superfluid state, the identities $\partial_{\mu}\mathbf{q}_{s}=\tfrac{1}{2}(\partial_{\mu}\phi_{s})\bm{\xi}_{3}$ and $\partial_{\mu}\bm{\xi}_{3}=-\tfrac{1}{2}(\partial_{\mu}\phi_{s})\mathbf{q}_{s}$ hold. Therefore, $\partial_{\mu}^{2}\mathbf{q}_{s}=\tfrac{1}{2}(\partial_{\mu}^{2}\phi_{s})\bm{\xi}_{3}-\tfrac{1}{4}(\partial_{\mu}\phi_{s})^{2}\mathbf{q}_{s}$ and $\bm{\omega}=(\partial_{t}\phi_{s})\bm{n}$, so that
\begin{align}
(\partial_{t}^{2}-v_{m}^{2}\nabla^{2})\mathbf{q}_{s}&=\tfrac{1}{2}\left(\partial_{t}^{2}\phi_{s}-v_{m}^{2}\partial_{x}^{2}\phi_{s}\right)\bm{\xi}_{3}-\tfrac{1}{4}\left[(\partial_{t}\phi_{s})^{2}-v_{m}^{2}(\partial_{x}\phi_{s})^{2}\right]\mathbf{q}_{s},\\
\lambda(\mathbf{q}_{s})&=\tfrac{1}{4}\left[(\partial_{t}\phi_{s})^{2}-v_{m}^{2}(\partial_{x}\phi_{s})^{2}\right]+\tfrac{\gamma}{2} B\partial_{t}\phi_{s},\\
\mathbf{B}^{\star}\wedge\partial_{t}\mathbf{q}&=\tfrac{1}{2}(B\partial_{t}\phi_{s})\mathbf{q}_{s},\\
\partial_{t}\mathbf{B}^{\star}\wedge\mathbf{q}&=-(\partial_{t}B)\bm{\xi}_{3}.
\end{align}
With account of all these identities Eq.~\eqref{EoM_q} becomes
\begin{equation}
\label{eq3}
\left[\partial_{t}^{2}\phi_{s}+\frac{1}{T}\partial_{t}\phi_{s}-v_{m}^{2}\partial_{x}^{2}\phi_{s}+\gamma\partial_{t}B\right]\bm{\xi}_{3}=\bm{0},
\end{equation}
which, in turn, leads to Eq.~\eqref{EoM_phi} since $\bm{\xi}_{3}\neq\bm{0}$. The separation of variables $\phi_{s}(x,t)=X(x)+\tau(t)$ yields, for a uniform external magnetic field, the system of equations
\begin{align}
\label{eq4}
\tau''(t)+\frac{1}{T}\tau'(t)+\gamma\partial_{t}B&=\mu,\\
\label{eq5}
X''(x)&=\frac{\mu}{v_{m}^{2}},
\end{align}
where $\mu$ is a constant to be determined. The solutions to these ordinary differential equations read 
\begin{align}
\label{eq6}
\tau(t)&=\tau_{0}+\mu T t-\gamma e^{-t/T}\int_{-\infty}^{t}B(t')e^{t'/T}dt',\\
\label{eq7}
X(x)&=X_{0}+kx+\frac{1}{2}\frac{\mu}{v_{m}^{2}}x^{2},
\end{align}
with $X_{0}$, $\tau_{0}$ and $k$ being constants to be determined by imposing boundary conditions. The latter, as discussed in Refs.~\onlinecite{Tserkovnyak-PRB2017} and~\onlinecite{Ochoa-PRB2018}, are given by $-\mathcal{A}\,\hat{n}_{\xi}\cdot\vec{\bm{\Omega}}|_{\xi}=\frac{g_{\xi}}{4\pi}\left[\bm{\mu}_{\xi}-\hbar\bm{\omega}_{\xi}\right]$, with $\xi=L,R$ indicating the terminals of the device. For the lateral configuration depicted in Fig.~\ref{Fig1} of the main text, we have $\hat{n}_{L}=-\hat{n}_{R}=\hat{x}$. By accounting for the parametrization $w=\cos(\phi/2)$, $\bm{v}=\sin(\phi/2)\bm{n}$ of the order parameter, Eqs.~\eqref{BC1} and \eqref{BC2} become 
\begin{eqnarray}
\label{BC1p}
\mathcal{A}\,\partial_{x}\phi\left(-\tfrac{L}{2},t\right)&=&\frac{g_{L}}{4\pi}\left[-\mu_{L}+\hbar\partial_{t}\phi\left(-\tfrac{L}{2},t\right)\right],\\
\label{BC2p}
\mathcal{A}\,\partial_{x}\phi\left(\tfrac{L}{2},t\right)&=&\frac{g_{R}}{4\pi}\left[\mu_{R}-\hbar\partial_{t}\phi\left(\tfrac{L}{2},t\right)\right].
\end{eqnarray}
Next we consider the phase-coherent precessional ansatz $\phi_{s}$, for which $\partial_{x}\phi_{s}=k+\mu x/v_{m}^{2}$ and $\partial_{t}\phi_{s}=\mu T$ under constant magnetic field. As a result, we need to solve the following linear system of equations for $k$ and $\mu$:
\begin{eqnarray}
\label{eq8}
\mathcal{A}\left[k+\frac{\mu}{v_{m}^{2}}\left(-\frac{L}{2}\right)\right]&=&\frac{g_{L}}{4\pi}\left[-\mu_{L}+\hbar\mu T\right],\\
\label{eq9}
\mathcal{A}\left[k+\frac{\mu}{v_{m}^{2}}\left(\frac{L}{2}\right)\right]&=&\frac{g_{R}}{4\pi}\left[\mu_{R}-\hbar\mu T\right],
\end{eqnarray}
from which the expressions \eqref{k} and \eqref{omega} ($\omega=\mu T$) follow.

\section{Derivation of equations (11) and (12)}

Equations~\eqref{EoM_l} and \eqref{EoM_t} are derived by incorporating the parametrization~\eqref{op_fluct} into Eq.~\eqref{EoM_q} and expanding up to first order in the fluctuation fields. The following intermediate results have been used:
\begin{align}
& 1)\hspace{0.1cm}\partial_{\mu}^{2}\mathbf{q}\simeq-\left[\tfrac{1}{4}(\partial_{\mu}\phi_{s})^{2}+\partial_{\mu}\Psi_{l}\partial_{\mu}\phi_{s}+\tfrac{1}{2}\Psi_{l}\partial_{\mu}^{2}\phi_{s}\right]\mathbf{q}_{s}+\hat{\bm{e}}_{+}\partial_{\mu}^{2}\Psi_{t}+\hat{\bm{e}}_{-}\partial_{\mu}^{2}\Psi_{t}^{*}+\left[\partial_{\mu}^{2}\Psi_{l}+\tfrac{1}{2}\partial_{\mu}^{2}\phi_{s}-\tfrac{1}{4}\Psi_{l}(\partial_{\mu}\phi_{s})^{2}\right]\bm{\xi}_{3},\\
& 2)\hspace{0.1cm}\partial_{\mu}\mathbf{q}\simeq-\tfrac{1}{2}\Psi_{l}\partial_{\mu}\phi_{s}\mathbf{q}_{s}+\hat{\bm{e}}_{+}\partial_{\mu}\Psi_{t}+\hat{\bm{e}}_{-}\partial_{\mu}\Psi_{t}^{*}+\left[\partial_{\mu}\Psi_{l}+\tfrac{1}{2}\partial_{\mu}\phi_{s}\right]\bm{\xi}_{3},\\
& 3)\hspace{0.1cm}\lambda(\mathbf{q})\simeq\left[\tfrac{1}{4}(\partial_{t}\phi_{s})^{2}-\tfrac{1}{4}v_{m}^{2}(\nabla\phi_{s})^{2}+\tfrac{\gamma}{2}\partial_{t}\phi_{s}B\right]+\left[\partial_{t}\Psi_{l}\partial_{t}\phi_{s}-v_{m}^{2}\nabla\Psi_{l}\cdot\nabla\phi_{s}+\gamma\partial_{t}\Psi_{l}B\right],\\
& 4)\hspace{0.1cm}\mathbf{B}^{\star}\wedge\partial_{t}\mathbf{q}\simeq-\tfrac{1}{2}(\Psi_{l}\partial_{t}\phi_{s})\mathbf{B}^{\star}\wedge\mathbf{q}_{s}+\partial_{t}\Psi_{t}\mathbf{B}^{\star}\wedge\hat{\bm{e}}_{+}+\partial_{t}\Psi_{t}^{*}\mathbf{B}^{\star}\wedge\hat{\bm{e}}_{-}+(\partial_{t}\Psi_{l}+\tfrac{1}{2}\partial_{t}\phi_{s})\mathbf{B}^{\star}\wedge\bm{\xi}_{3},\\
& 5)\hspace{0.1cm}\partial_{t}\mathbf{B}^{\star}\wedge\mathbf{q}\simeq \partial_{t}\mathbf{B}^{\star}\wedge\mathbf{q}_{s}+\Psi_{t}\partial_{t}\mathbf{B}^{\star}\wedge\hat{\bm{e}}_{+}+\Psi_{t}^{*}\partial_{t}\mathbf{B}^{\star}\wedge\hat{\bm{e}}_{-}+\Psi_{l}\partial_{t}\mathbf{B}^{\star}\wedge\bm{\xi}_{3},\\
& 6)\hspace{0.1cm}\mathbf{B}^{\star}\wedge\mathbf{q}_{s}=-B\bm{\xi}_{3},\hspace{0.1cm} \hspace{0.1cm}\mathbf{B}^{\star}\wedge\bm{\xi}_{3}=B\mathbf{q}_{s},\hspace{0.1cm}\mathbf{B}^{\star}\wedge\hat{\bm{e}}_{\pm}=\pm i B\hat{\bm{e}}_{\pm}.
\end{align} 

\section{Stability analysis}
As discussed in the main text, the phase coherent precessional state \eqref{SS_sol} is robust against longitudinal fluctuations $\Psi_{l}$. Furthermore, transverse fluctuations proliferate when the exponent $-\tfrac{1}{T}\pm\textrm{Re}(Z)$ is positive, which determines the instability threshold for the spin superfluid solution. Since $\textrm{Re}(Z)=r^{1/2}\cos(\theta/2)$, with $re^{i\theta}\equiv Z^{2}=\left[\tfrac{1}{T^{2}}-\gamma^{2}B^{2}-\omega^{2}-v_{m}^{2}(4q^{2}-k^{2})-2\gamma\omega B\right]-i\left[\tfrac{2\gamma B}{T}\right]$, this threshold can be recast as
\begin{equation}
\label{eq10}
r+r\cos\theta\geq\frac{2}{T^{2}},
\end{equation}
which, after some algebra, yields the inequality $v_{m}^{2}k^{2}\geq\omega^{2}+4v_{m}^{2}q^{2}+2\gamma\omega B$. The highest magnetic field satisfying the latter inequality occurs for $q=0$ (see the first paragraph of the Discussion section in the man text, however), which leads to Eq.~\eqref{critical_B} in the main text.

\end{document}